\documentclass{v16nufact}

\def\be{\begin{equation}}
\def\ee{\end{equation}}
\def\bea{\begin{eqnarray}}
\def\eea{\end{eqnarray}}

\usepackage{cite}
\newcommand{\Photo}{\includegraphics[height=35mm]{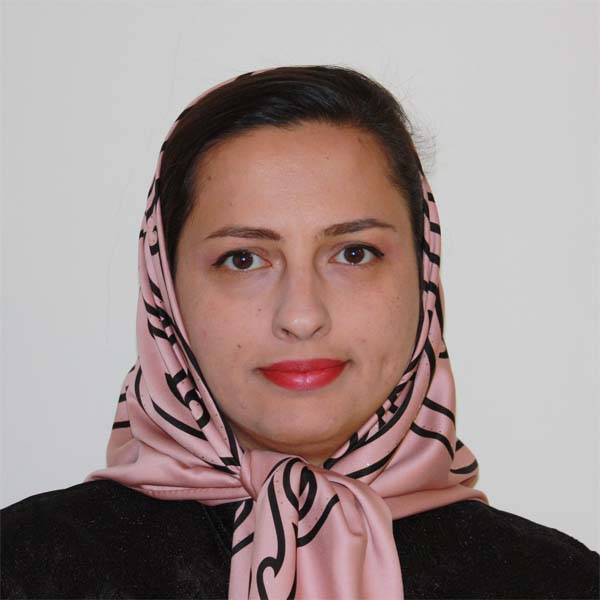}}

\begin{document}
\vspace*{4cm}
\title{VIABLE MODELS FOR LARGE NON-STANDARD NEUTRINO INTERACTION}

\author{ Yasaman Farzan }

\address{
School of physics, Institute for
      Research in Fundamental Sciences (IPM),\\
      P.O. Box 19395-5531, Tehran, Iran}

\maketitle\abstracts{Thanks to high precision  long baseline neutrino experiments such as NO$\nu$A and DUNE, possible effects of Non-Standard neutrino Interactions (NSI) on neutrino oscillation data have received renewed interest in the last two years. It is however challenging to build models that can give rise to NSI with sizeable couplings  discernible at neutrino oscillation experiments without violating the various existing experimental bounds. We introduce two viable models that lead to neutral current NSI with sizable couplings. Both models are based on a new $U(1)^\prime$ gauge symmetry with $Z^\prime$ gauge boson of mass $O(10~{\rm MeV})$. We will highlight the common phenomenological features of these models and suggest ways to test them.
}

\section{Introduction}

The state-of-the-art long baseline neutrino experiments, such as current  NO$\nu$A  or proposed DUNE experiment, are ushering in precision era in measurement of neutrino
oscillation parameters and promise to determine the yet-unknown neutrino oscillation parameters:  $\delta_{CP}$, ${\rm sign}(\Delta m_{31}^2)$ and the octant of $\theta_{23}$. The first natural question that rises is that whether we know the neutrino interactions well enough
to be able  to resolve the subdominant effects in order to extract the yet-unknown neutrino oscillation parameters.
It is well-known that neutral current NSI between neutrino and matter fields ({\it i.e.,} electrons, up- and/or down-quark) can change the so-called effects in neutrino oscillation in a medium  \cite{Friedland:2012tq,Masud:2015xva,deGouvea:2015ndi,Coloma:2015kiu,Masud:2016bvp,Masud:2016gcl,Bakhti:2016gic}.
It has been demonstrated that even maximal CP-violation ($\delta=270^\circ$) can be faked by NSI in NO$\nu$A and DUNE experiments despite conservation of CP at lepton sector \cite{Forero:2016cmb}.
Moreover it has been shown that the introduction of NSI can induce degeneracies in determination of the octant of $\theta_{23}$ \cite{Agarwalla:2016fkh} \footnote{It is shown in \cite{Bakhti:2016prn} that set-ups such as the long baseline MOMENT experiment \cite{Cao:2014bea} proposed to measure $\delta_{CP}$ are ideal to resolve such degeneracy simply because due to shorter baseline and lower beam energy relative to NO$\nu$A and DUNE is not sensitive to matter effect (neither standard nor NSI).}.
Moreover, it has been shown that turning on neutral current NSI, along with standard LMA solution to the solar neutrino anomaly with $\theta_{12}<\pi/4$, another solution known as LMA-Dark solution with $\theta_{12} >\pi/4$ appears. Surprisingly  this new solution can provide even a better fit to solar neutrino date \cite{Gonzalez-Garcia:2013usa,Miranda:2004nb,Escrihuela:2009up}. The LMA-Dark solution requires the effective NSI coupling to be comparable in magnitude with standard Fermi coupling $G_F$, which of course from model building point of view is very challenging.

The neutral current NSI in question can be parameterized as
\begin{equation}
\L_{ \rm NSI} = -2\sqrt2 G_F \,\epsilon_{\alpha\beta}^{f X}\, (\bar\nu_\alpha \gamma^\mu P_L \nu_\beta)(\bar f \gamma_\mu P_X f)\,,
\label{eq:NSI}
\end{equation}
where $P_{R/L} \equiv (1\pm \gamma_5)/2$ is the chirality projection operator, index $X$ may denote $L$ or $R$, $f\in\{e,u,d\}$ specifies the matter particles, and $\alpha, \beta\in \{e,\mu,\tau\}$ denote the neutrino flavor.
The combinations of $\epsilon$ that are relevant for neutrino oscillation in matter are  $\epsilon_{\alpha \beta}\equiv\sum_f (n_f/n_e)(\epsilon_{\alpha \beta}^{fL}+\epsilon_{\alpha \beta}^{fR})$. To be more precise, the neutrino oscillation pattern is sensitive only to  $\epsilon_{\alpha \beta}$ with $\alpha\ne \beta$ and splitting of diagonal elements
$(\epsilon_{\alpha \alpha}-\epsilon_{\beta \beta})$. This can be understood because adding or subtracting a matrix proportional to $1_{3\times 3}$ to the Hamiltonian governing the neutrino propagation does not change the neutrino oscillation pattern. In the limit $\epsilon \to 0$, we recover the standard case without any new effect on neutrino oscillation.

The second natural question that arises is the following: Is there a viable model that can give rise to $\epsilon_{\alpha \beta}$ large enough to lead to discernable effects at neutrino  oscillation experiments?
Being inspired by the Fermi effective Lagrangian, the first idea for model building that comes to mind is that the effective Lagrangian in Eq. (\ref{eq:NSI})
originates from integrating out a heavy state, $X$. If we demand the mass of the heavy state, $m_X$, to be large enough to avoid direct production at colliders, $\epsilon_{\alpha \beta}$ should be suppressed by $G_F^{-1} m_X^{-2}$. Instead of taking $m_X\gg m_W$, Refs \cite{Farzan:2015doa,Farzan:2015hkd,Farzan:2016wym} suggested to identify $X$ with a new $U(1)^\prime$ gauge boson $Z^\prime$ with mass $m_{Z^\prime} \sim O(10~{\rm MeV})$.  As far as neutrino oscillation in medium is concerned, we can still use the effective four-Fermi interaction  in Eq. (\ref{eq:NSI}) even if the energy of the neutrino beam is larger than the mass of the intermediate particle. This  is because for neutrino propagation in matter only forward scattering with zero energy momentum transfer is relevant. However, at neutrino scattering experiments such as CCFR \cite{Mishra:1991bv}, CHARM \cite{Geiregat:1990gz} and NuTeV \cite{Davidson:2003ha}, the amplitude of new contribution relative to standard model contribution will be
suppressed by a factor of $\epsilon_{\alpha \beta}^f m_{Z^\prime}^2/q^2$ where $q$ is the typical energy -momentum transfer in the scattering for $m_{Z^\prime} \ll 1$ GeV. Relevant bounds from these experiments can be therefore relaxed.
Throughout this letter, we set $\epsilon_{\alpha \beta}^{eL}=\epsilon_{\alpha \beta}^{eR}=0$ in order not to affect the solar neutrino flux at the Borexino and SNO experiments. Moreover, by setting  $\epsilon^{d L}_{\alpha \beta}=\epsilon^{ d R}_{\alpha \beta}$ and $\epsilon^{u  L}_{\alpha \beta}=\epsilon^{u  R}_{\alpha \beta}$, the measurement of neutral current interaction rate of solar neutrino flux at SNO experiment (being a Gamow-Teller $\nu +D\to \nu+p+n$ process) will not be affected. Because the coupling to quarks is taken to be non-chiral, we simply drop the chirality projection index $L$ and $R$.

The present letter is organized as follows. In sect. \ref{diag}, we briefly review the model in Ref \cite{Farzan:2015doa} which provides a basis for LMA-Dark solution. In sect. \ref{LFV}, we review the model in Ref. \cite{Farzan:2016wym} which can give rise to flavor diagonal ({\it e.g.,} LMA-Dark solution) as well as flavor off-diagonal structure for $\epsilon_{\alpha \beta}$. In sect. \ref{common}, we outline some of observational predictions common between these two models. Conclusions are summarized in sect. \ref{con}.

\section{A model for LMA-Dark}\label{diag} In this section, we briefly review the model introduced in
Ref \cite{Farzan:2015doa} to embed the LMA-Dark solution. This solution requires
\begin{equation}
\epsilon_{\alpha \beta}^q|_{\alpha \ne \beta}, \epsilon_{\mu \mu}^q-\epsilon_{\tau\tau}^q \ll \epsilon^q_{\mu \mu}-\epsilon^q_{ee}\simeq \epsilon^q_{\tau \tau}-\epsilon^q_{ee}\sim 1~~{\rm  where}  ~~q=u~{\rm and/or} ~d \label{pattern} .\end{equation}
Such a pattern can be obtained by gauging a combination $L_\mu +L_\tau+b B_1$ where $L_\mu$, $L_\tau$ and $B_1$ are respectively lepton numbers of second and third generations and Baryon number of first generation. $b$ is a positive number which for simplicity is set equal to 1 in Ref \cite{Farzan:2015doa}. We then obtain $\epsilon_{ee}^q=\epsilon_{\alpha \beta}^q|_{\alpha \ne \beta}=0$ and
\begin{equation}
\epsilon_{\mu \mu}^q =\epsilon_{\tau \tau}^q=\frac{b g^{\prime 2}}{3 \sqrt{2} G_F m_{Z^\prime}^2 } \end{equation} where $g^\prime$ is the new gauge coupling to obtain $\epsilon_{\mu \mu}^q =\epsilon_{\tau \tau}^q\sim 1$, we need \begin{equation} g^\prime\sim 7 \times 10^{-5}\left(\frac{m_{Z^\prime}}{10~{\rm MeV}}\right) \left( \frac{1}{b}\right)^{1/2}.\end{equation}
Notice that along with $\nu_\mu$ and $\nu_\tau$, $\mu$ and $\tau$ will also obtain new interaction with the same coupling. The value of $g^\prime$ required for the LMA-Dark solution is smaller than the present upper bounds from observations such as $(g-2)_\mu$ and unfortunately is too small to explain the famous $(g-2)_\mu$ anomaly \cite{Jegerlehner:2009ry}. Notice that the tree level electron has no new coupling so the restrictive bounds on new interactions of the electron can be easily avoided. For a more comprehensive discussion of the bounds see Ref \cite{Farzan:2015doa}.

To cancel gauge anomalies, Ref. \cite{Farzan:2015doa} has suggested to gauge the anomaly free combination $L_\mu +L_\tau+B_1+B_2 -4B_3$ where $B_2$ and $B_3$ are baryon numbers of second and third generations.  With this combination, the mixing between first and second generation of quarks as well as the mixing between the second and third generations of lepton can be readily obtained without breaking
the new gauge symmetry. However, to regenerate full CKM and PMNS mixing matrices, we need new scalars whose VEV break $U(1)^\prime$. The new Higgs doublet added to mix the third generation of quarks to the rest can be produced at the LHC. Ref.  \cite{Farzan:2015doa} suggests a mechanism to reduce its VEV below electroweak scale despite its mass being higher. Neutrinos obtain mass via type I seesaw mechanism after $U(1)^\prime$ symmetry breaking. The same scalars give mass to $Z^\prime$.
To obtain $m_{Z^\prime} \sim 10$ MeV for $g^\prime \sim 7 \times 10^{-5}$, the largest VEV of new scalars (which are charged under $U(1)^\prime$ but are singlets under the standard model gauge group) should be of order of TeV,

 \section{NSI through mixing between neutrino and a  Dirac sterile fermion}\label{LFV}
Building a model which gives rise to lepton flavor violating NSI ({\it i.e.,} $\epsilon_{\alpha \beta} \ne 0$ for $\alpha \ne \beta$) is more challenging.  Ref. \cite{Farzan:2015hkd} tries this by assigning opposite $U(1)^\prime$ charges to orthogonal combinations of $L_\alpha$ which are not aligned with mass eigenvectors. This way LFV NSI for neutrino can be obtained but there will be also similar couplings for charged leptons leading to fast $l_\alpha^-\to l_\beta^- Z^\prime$ unless coupling of $Z^\prime$ to leptons is smaller than $10^{-9}$. Ref. \cite{Farzan:2016wym} takes another approach. In this model leptons are not gauged under $U(1)^\prime$. Instead, a new Dirac fermion  denoted by $\Psi$ has been added which is singlet under standard model gauge symmetry but is charged under $U(1)^\prime$. Moreover a new Higgs doublet denoted by $H^\prime$ is added whose $U(1)^\prime$ charge is equal to that of $\Psi$. That is under $U(1)^\prime$, $\Psi \to e^{ig_{\Psi}} \Psi$ and $H^\prime \to e^{i g_{\Psi}} H^\prime$. As a result a Yukawa coupling of the following form can be written as
\begin{equation}
\ L = -\sum_\alpha y_\alpha \overline{L}_\alpha \tilde H' P_R \Psi + {\rm H.c.}
\label{eq:newyukawa}
\end{equation}
The VEV of $H^\prime$ parameterized as $\langle H^\prime \rangle=v \cos \beta /\sqrt{2}$ with $v=246$ GeV breaks both electroweak and $U(1)^\prime$ symmetries and induces mixing between neutrinos and $\Psi$ given by
\begin{equation}
\label{k-a}
\kappa_\alpha = \frac{y_\alpha\langle H'\rangle}{M_\Psi} = \frac{y_\alpha v \cos\beta}{\sqrt2 M_\Psi} \ .\end{equation}
Notice that because of this mixing, the PMNS matrix deviates from unitarity. There are bounds on the violation of the unitarity from the muon decay measurement and/or tests of lepton flavor universality \cite{Fernandez-Martinez:2016lgt}:
\be
\label{Di} |\kappa_e|^2<2.5 \times 10^{-3},  \ |\kappa_\mu|^2<4.4 \times 10^{-4},  \ {\rm and} \  |\kappa_\tau|^2<5.6 \times 10^{-3} \ {\rm at } \ 2\sigma.
\ee
Ref \cite{Fernandez-Martinez:2016lgt} derives stronger bounds from  $Br(\mu\to e \gamma)$ limit on the product $\kappa_e \kappa_\mu$; however, this bound does not apply to our case because $\Psi$ has a mass of few GeV and therefore GIM mechanism suppresses the contribution to $\mu \to e \gamma$. Through the mixing, $\Psi$ can be produced at the high energy neutrino scattering experiments such as NuTeV but since its main decay mode is invisible $\Psi \to Z^\prime \nu$ (and subsequently $Z^\prime \to \nu \bar{\nu}$), no significant bound can be set on the $\kappa_\alpha$ mixing from these experiments.

Through the $\kappa$ mixing neutrinos couple to $Z^\prime$ with a coupling
\begin{equation} g_\Psi \sum_{\alpha, \beta} \kappa_\alpha^* \kappa_\beta (\bar{\nu}_\alpha \gamma^\mu \nu_\beta) Z^\prime_\mu. \end{equation}
On the other hand, quarks under $U(1)^\prime$ transform as $q \to e^{i g_B/3} q$. \footnote{To cancel anomalies, it is suggested to add new generations of
leptons \cite{Farzan:2016wym}.}
We therefore obtain
\begin{equation}
\epsilon_{\alpha\beta}^{u}=\epsilon_{\alpha\beta}^{d} \simeq  \frac{g_B g_\Psi \kappa_\alpha^* \kappa_\beta}{6\sqrt2 G_F M_{Z'}^2}.\end{equation}
Notice that assigning opposite signs to $g_B$ and $g_\Psi$ and taking $(|g_B g_\Psi|)^{1/2} \sim 10^{-4} (m_{Z^\prime}/(10~{\rm MeV})$, we can reproduce the LMA-Dark solution. Moreover, if
$\Psi$ mixes with more than one generation, we can have lepton flavor violating NSI with $|\epsilon_{\alpha \beta}|=\sqrt{|\epsilon_{\alpha \alpha} \epsilon_{\beta \beta}|}$. When $y_\alpha y_\beta^*$ is complex, the off-diagonal elements of $\epsilon_{\alpha \beta}$ can be also complex, inducing new sources of CP-violation for neutrino oscillation.
\section{Observational effects}\label{common}
In these models, we have a light $O(10~{\rm MeV})$ new particle with couplings to both quarks and neutrinos of order of $\sim O(5\times 10^{-5} -10^{-4})$. Not surprisingly, we expect observable effects in a myriad experiments and observations. A comprehensive list of effects can be found in Ref \cite{Farzan:2015doa,Farzan:2016wym}. Here, we only emphasize on the effects that provide promising tests for the model(s). For example effects on big bang nucleosynthesis yield \cite{Kamada:2015era} $$m_{Z^\prime}> 5 ~{\rm MeV}.$$
Other important effects include (i) effects on duration  of neutrino emission from supernova type II; (ii) dip in the energy spectrum of high cosmic neutrinos and, (iii)
rates of interaction of solar neutrinos at direct DM search experiments.  Below we briefly review each effect one by one.  We should however first notice that for  $m_{Z^\prime}<m_\pi$, the dominant decay mode of $Z^\prime$ is decay to neutrinos.

i) We expect the $Z^\prime$ to be thermally produced in the supernova core via neutrino pair annihilation and decay back to neutrinos inside the core with decay length
$$c \tau =10^{-9}{\rm km} \left( \frac{g^\prime}{7 \times 10^{-5}}\right)^{-2} \left(\frac{T}{10~{\rm MeV}}\right) \left(\frac{10~{\rm MeV}}{m_{Z^\prime}}\right)^2.$$
This new interaction between neutrinos reduces the mean free path of neutrinos inside supernova core which in turn prolongs the duration of supernova neutrino emission. Ref. \cite{Kamada:2015era} estimates that within the parameter range of our interest,  the prolongation can be large enough to be resolved in the event of a galactic supernova detection. To quantitatively  derive the effect,
full simulation is required.

ii)  High energy cosmic neutrinos on their way to Earth can interact with background relic neutrinos. If the center of mass energy of the two neutrinos is equal to $m_{Z^\prime}$, $Z^\prime$ can be resonantly produced and decay back to a pair of neutrinos whose momenta are smaller than the momenta of the initial high energy  neutrinos. The dip is expected to be located at $E_\nu \sim \sqrt{m_{Z^\prime}^2/T_\nu}$ where $T_\nu \sim 10^{-4}$ eV is the temperature of background neutrinos so far $m_{Z^\prime}\sim 10$ MeV, we expect the dip to lie around 500~TeV-1~PeV. As shown in  \cite{Kamada:2015era},  for the values of gauge couplings of our interest, the optical depth can be larger than one making the dip discernable. See also Refs \cite{Hooper:2007jr,Ioka:2014kca,Ng:2014pca,DiFranzo:2015qea}.
In fact, there is already a hint for such dip in ICECUBE data but confirmation requires more data points.

iii) As is well-known the scattering of solar neutrino flux at the experiments designed to directly detect dark  matter can provide a background. Ref. \cite{Cerdeno:2016sfi} has shown that
the measurement of the interaction rate of the solar neutrino flux at these experiments with both electrons and nuclei can probe new gauge interactions of neutrinos. The best present bounds comes from CDMLite experiment \cite{Agnese:2015nto}: $\sqrt{g_B g_\nu}\stackrel{<}{\sim} 5 \times 10^{-5}$ for $m_{Z^\prime} \sim 5$ MeV. As shown in Ref \cite{Cerdeno:2016sfi}, this bound already rules out a part of parameter space relevant for LMA-Dark solution.
Future bounds from LUX-ZEPLIN \cite{Akerib:2015cja} and SuperCDMS \cite{Brink:2012zza} can fully probe the parameter space that we are interested in.
\section{Summary}\label{con}
We have presented two models that can give rise to  neutral current NSI for neutrinos large enough to be discernable at the  neutrino
oscillation experiments. Both models are based on a new $U(1)^\prime$ gauge interaction with gauge boson of mass $\sim 10$~MeV and coupling to neutrinos and first generation quarks of order of $5\times10^{-5}-10^{-4}$. The models can be tested by various observations including studying the effects on prolongation of the duration of neutrino emission from supernova type II, searching for a dip at $E_\nu \sim 400~{\rm TeV}-1~{\rm PeV}$ in the energy spectrum of cosmic neutrinos and measurements of the coherent interaction rates of solar neutrino flux off nuclei in the future direct dark matter search experiments.

\section*{Acknowledgments}
The author is grateful to organizers of NuFact2016 and the staff of ICISE in Vietnam where this work was presented.
She is also grateful to Julian Heeck and Enrique Fernandez-Martinez for useful discussions.
This project has received funding from the European Union's Horizon 2020 research and innovation programme under the Marie Sk\l{}odowska-Curie grant agreement No.~674896 and No.~690575.
YF is also grateful to the ICTP associate office and Iran National Science Foundation (INSF) for partial financial support under contract 94/saad/43287.

\section*{References}

%\bibliographystyle{utcaps_mod}
%\bibliography{references}

\begin{thebibliography}{99}
\bibitem{Friedland:2012tq}
  A.~Friedland and I.~M.~Shoemaker,
  ``Searching for Novel Neutrino Interactions at NOvA and Beyond in Light of Large $\theta_{13}$,''
  arXiv:1207.6642 [hep-ph].
  %%CITATION = ARXIV:1207.6642;%%
\bibitem{Masud:2015xva}
M.~Masud, A.~Chatterjee and P.~Mehta,
  ``Probing CP violation signal at DUNE in presence of non-standard neutrino interactions,''
  J.\ Phys.\ G {\bf 43} (2016) no.9,  095005
  doi:10.1088/0954-3899/43/9/095005/meta, 10.1088/0954-3899/43/9/095005
  [arXiv:1510.08261 [hep-ph]].
  %%CITATION = doi:10.1088/0954-3899/43/9/095005/meta, 10.1088/0954-3899/43/9/095005;%%
\bibitem{deGouvea:2015ndi}
 A.~de Gouvêa and K.~J.~Kelly,
  ``Non-standard Neutrino Interactions at DUNE,''
  Nucl.\ Phys.\ B {\bf 908} (2016) 318
  doi:10.1016/j.nuclphysb.2016.03.013
  [arXiv:1511.05562 [hep-ph]].
  %%CITATION = doi:10.1016/j.nuclphysb.2016.03.013;%%


\bibitem{Coloma:2015kiu}
P.~Coloma,
  ``Non-Standard Interactions in propagation at the Deep Underground Neutrino Experiment,''
  JHEP {\bf 1603} (2016) 016
  doi:10.1007/JHEP03(2016)016
  [arXiv:1511.06357 [hep-ph]].
  %%CITATION = doi:10.1007/JHEP03(2016)016;%%

\bibitem{Masud:2016bvp}
M.~Masud and P.~Mehta,
  ``Nonstandard interactions spoiling the CP violation sensitivity at DUNE and other long baseline experiments,''
  Phys.\ Rev.\ D {\bf 94} (2016) 013014
  doi:10.1103/PhysRevD.94.013014
  [arXiv:1603.01380 [hep-ph]].
  %%CITATION = doi:10.1103/PhysRevD.94.013014;%%

\bibitem{Masud:2016gcl}
 M.~Masud and P.~Mehta,
  ``Nonstandard interactions and resolving the ordering of neutrino masses at DUNE and other long baseline experiments,''
  Phys.\ Rev.\ D {\bf 94} (2016) no.5,  053007
  doi:10.1103/PhysRevD.94.053007
  [arXiv:1606.05662 [hep-ph]].
  %%CITATION = doi:10.1103/PhysRevD.94.053007;%%

\bibitem{Bakhti:2016gic}
 P.~Bakhti and A.~N.~Khan,
  ``Sensitivities to charged-current nonstandard neutrino interactions at DUNE,''
  arXiv:1607.00065 [hep-ph].
  %%CITATION = ARXIV:1607.00065;%%
\bibitem{Forero:2016cmb}
  D.~V.~Forero and P.~Huber,
  ``Hints for leptonic CP violation or New Physics?,''
  Phys.\ Rev.\ Lett.\  {\bf 117} (2016) no.3,  031801
  doi:10.1103/PhysRevLett.117.031801
  [arXiv:1601.03736 [hep-ph]].
  %%CITATION = doi:10.1103/PhysRevLett.117.031801;%%
\bibitem{Agarwalla:2016fkh}
 S.~K.~Agarwalla, S.~S.~Chatterjee and A.~Palazzo,
  ``Degeneracy between $\theta_{23}$ octant and neutrino non-standard interactions at DUNE,''
  Phys.\ Lett.\ B {\bf 762} (2016) 64
  doi:10.1016/j.physletb.2016.09.020
  [arXiv:1607.01745 [hep-ph]].
  %%CITATION = doi:10.1016/j.physletb.2016.09.020;%%


 \bibitem{Bakhti:2016prn}
 P.~Bakhti and Y.~Farzan,
  ``CP-Violation and Non-Standard Interactions at the MOMENT,''
  JHEP {\bf 1607} (2016) 109
  doi:10.1007/JHEP07(2016)109
  [arXiv:1602.07099 [hep-ph]].
  %%CITATION = doi:10.1007/JHEP07(2016)109;%%

 \bibitem{Cao:2014bea}
 J.~Cao {\it et al.},
  ``Muon-decay medium-baseline neutrino beam facility,''
  Phys.\ Rev.\ ST Accel.\ Beams {\bf 17} (2014) 090101
  doi:10.1103/PhysRevSTAB.17.090101
  [arXiv:1401.8125 [physics.acc-ph]].
  %%CITATION = doi:10.1103/PhysRevSTAB.17.090101;%%

\bibitem{Gonzalez-Garcia:2013usa}
M.~C.~Gonzalez-Garcia and M.~Maltoni,
  ``Determination of matter potential from global analysis of neutrino oscillation data,''
  JHEP {\bf 1309} (2013) 152
  doi:10.1007/JHEP09(2013)152
  [arXiv:1307.3092 [hep-ph]].
  %%CITATION = doi:10.1007/JHEP09(2013)152;%%
\bibitem{Miranda:2004nb}
O.~G.~Miranda, M.~A.~Tortola and J.~W.~F.~Valle,
  ``Are solar neutrino oscillations robust?,''
  JHEP {\bf 0610} (2006) 008
  doi:10.1088/1126-6708/2006/10/008
  [hep-ph/0406280].
  %%CITATION = doi:10.1088/1126-6708/2006/10/008;%%
\bibitem{Escrihuela:2009up}
 F.~J.~Escrihuela, O.~G.~Miranda, M.~A.~Tortola and J.~W.~F.~Valle,
  ``Constraining nonstandard neutrino-quark interactions with solar, reactor and accelerator data,''
  Phys.\ Rev.\ D {\bf 80} (2009) 105009
   Erratum: [Phys.\ Rev.\ D {\bf 80} (2009) 129908]
  doi:10.1103/PhysRevD.80.129908, 10.1103/PhysRevD.80.105009
  [arXiv:0907.2630 [hep-ph]].
  %%CITATION = doi:10.1103/PhysRevD.80.129908, 10.1103/PhysRevD.80.105009;%%
\bibitem{Farzan:2015doa}
 Y.~Farzan,
  ``A model for large non-standard interactions of neutrinos leading to the LMA-Dark solution,''
  Phys.\ Lett.\ B {\bf 748} (2015) 311
  doi:10.1016/j.physletb.2015.07.015
  [arXiv:1505.06906 [hep-ph]].
  %%CITATION = doi:10.1016/j.physletb.2015.07.015;%%
\bibitem{Farzan:2015hkd}
 Y.~Farzan and I.~M.~Shoemaker,
  ``Lepton Flavor Violating Non-Standard Interactions via Light Mediators,''
  JHEP {\bf 1607} (2016) 033
  doi:10.1007/JHEP07(2016)033
  [arXiv:1512.09147 [hep-ph]].
  %%CITATION = doi:10.1007/JHEP07(2016)033;%%

\bibitem{Farzan:2016wym}
Y.~Farzan and J.~Heeck,
  ``Neutrinophilic nonstandard interactions,''
  Phys.\ Rev.\ D {\bf 94} (2016) no.5,  053010
  doi:10.1103/PhysRevD.94.053010
  [arXiv:1607.07616 [hep-ph]].
  %%CITATION = doi:10.1103/PhysRevD.94.053010;%%
\bibitem{Mishra:1991bv}
S.~R.~Mishra {\it et al.} [CCFR Collaboration],
  ``Neutrino tridents and W Z interference,''
  Phys.\ Rev.\ Lett.\  {\bf 66} (1991) 3117.
  doi:10.1103/PhysRevLett.66.3117
  %%CITATION = doi:10.1103/PhysRevLett.66.3117;%%
\bibitem{Geiregat:1990gz}
D.~Geiregat {\it et al.} [CHARM-II Collaboration],
 ``First observation of neutrino trident production,''
  Phys.\ Lett.\ B {\bf 245} (1990) 271.
  doi:10.1016/0370-2693(90)90146-W
  %%CITATION = doi:10.1016/0370-2693(90)90146-W;%%
\bibitem{Davidson:2003ha}
 S.~Davidson, C.~Pena-Garay, N.~Rius and A.~Santamaria,
``Present and future bounds on nonstandard neutrino interactions,''
  JHEP {\bf 0303} (2003) 011
  doi:10.1088/1126-6708/2003/03/011
  [hep-ph/0302093].
  %%CITATION = doi:10.1088/1126-6708/2003/03/011;%%
\bibitem{Jegerlehner:2009ry}
F.~Jegerlehner and A.~Nyffeler,
  ``The Muon g-2,''
  Phys.\ Rept.\  {\bf 477} (2009) 1
  doi:10.1016/j.physrep.2009.04.003
  [arXiv:0902.3360 [hep-ph]].
  %%CITATION = doi:10.1016/j.physrep.2009.04.003;%%

\bibitem{Fernandez-Martinez:2016lgt}
E.~Fernandez-Martinez, J.~Hernandez-Garcia and J.~Lopez-Pavon,
 ``Global constraints on heavy neutrino mixing,''
  JHEP {\bf 1608} (2016) 033
  doi:10.1007/JHEP08(2016)033
  [arXiv:1605.08774 [hep-ph]].
  %%CITATION = doi:10.1007/JHEP08(2016)033;%%
\bibitem{Kamada:2015era}
A.~Kamada and H.~B.~Yu,
 ``Coherent Propagation of PeV Neutrinos and the Dip in the Neutrino Spectrum at IceCube,''
  Phys.\ Rev.\ D {\bf 92} (2015) no.11,  113004
  doi:10.1103/PhysRevD.92.113004
  [arXiv:1504.00711 [hep-ph]].
  %%CITATION = doi:10.1103/PhysRevD.92.113004;%%
\bibitem{Hooper:2007jr}
  D.~Hooper,
  ``Detecting MeV gauge bosons with high-energy neutrino telescopes,''
  Phys.\ Rev.\ D {\bf 75} (2007) 123001
  doi:10.1103/PhysRevD.75.123001
  [hep-ph/0701194].
  %%CITATION = doi:10.1103/PhysRevD.75.123001;%%
\bibitem{Ioka:2014kca}
K.~Ioka and K.~Murase,
  ``IceCube PeV–EeV neutrinos and secret interactions of neutrinos,''
  PTEP {\bf 2014} (2014) no.6,  061E01
  doi:10.1093/ptep/ptu090
  [arXiv:1404.2279 [astro-ph.HE]].
  %%CITATION = doi:10.1093/ptep/ptu090;%%
 \bibitem{Ng:2014pca}
 K.~C.~Y.~Ng and J.~F.~Beacom,
  ``Cosmic neutrino cascades from secret neutrino interactions,''
  Phys.\ Rev.\ D {\bf 90} (2014) no.6,  065035
   Erratum: [Phys.\ Rev.\ D {\bf 90} (2014) no.8,  089904]
  doi:10.1103/PhysRevD.90.065035, 10.1103/PhysRevD.90.089904
  [arXiv:1404.2288 [astro-ph.HE]].
  %%CITATION = doi:10.1103/PhysRevD.90.065035, 10.1103/PhysRevD.90.089904;%%
 \bibitem{DiFranzo:2015qea}
A.~DiFranzo and D.~Hooper,
  ``Searching for MeV-Scale Gauge Bosons with IceCube,''
  Phys.\ Rev.\ D {\bf 92} (2015) no.9,  095007
  doi:10.1103/PhysRevD.92.095007
  [arXiv:1507.03015 [hep-ph]].
  %%CITATION = doi:10.1103/PhysRevD.92.095007;%%
\bibitem{Cerdeno:2016sfi}
 D.~G.~Cerdeño, M.~Fairbairn, T.~Jubb, P.~A.~N.~Machado, A.~C.~Vincent and C.~Bœhm,
 ``Physics from solar neutrinos in dark matter direct detection experiments,''
  JHEP {\bf 1605} (2016) 118
   Erratum: [JHEP {\bf 1609} (2016) 048]
  doi:10.1007/JHEP09(2016)048, 10.1007/JHEP05(2016)118
  [arXiv:1604.01025 [hep-ph]].
  %%CITATION = doi:10.1007/JHEP09(2016)048, 10.1007/JHEP05(2016)118;%%
\bibitem{Agnese:2015nto}
R.~Agnese {\it et al.} [SuperCDMS Collaboration],
 ``New Results from the Search for Low-Mass Weakly Interacting Massive Particles with the CDMS Low Ionization Threshold Experiment,''
  Phys.\ Rev.\ Lett.\  {\bf 116} (2016) no.7,  071301
  doi:10.1103/PhysRevLett.116.071301
  [arXiv:1509.02448 [astro-ph.CO]].
  %%CITATION = doi:10.1103/PhysRevLett.116.071301;%%
\bibitem{Akerib:2015cja}
 D.~S.~Akerib {\it et al.} [LZ Collaboration],
  ``LUX-ZEPLIN (LZ) Conceptual Design Report,''
  arXiv:1509.02910 [physics.ins-det].
  %%CITATION = ARXIV:1509.02910;%%
\bibitem{Brink:2012zza}
P.~L.~Brink [SuperCDMS Collaboration],
  ``Conceptual Design for SuperCDMS SNOLAB,''
  J.\ Low.\ Temp.\ Phys.\  {\bf 167} (2012) 1093.
  doi:10.1007/s10909-011-0440-3
  %%CITATION = doi:10.1007/s10909-011-0440-3;%%


%\bibitem{ja}C Jarlskog in {\em CP Violation}, ed. C Jarlskog
%(World Scientific, Singapore, 1988).

%\bibitem{ma}L. Maiani, \Journal{\PLB}{62}{183}{1976}.

%\bibitem{bu}J.D. Bjorken and I. Dunietz, \Journal{\PRD}{36}{2109}{1987}.

%\bibitem{bd}C.D. Buchanan {\it et al}, \Journal{\PRD}{45}{4088}{1992}.

\end{thebibliography}

\end{document}